\documentstyle[amssymb,12pt]{amsart}
\newtheorem{thm}{Theorem}
\newtheorem{prop}[thm]{Proposition}
\theoremstyle{remark}
\newtheorem{rem}{Remark}

\begin{document}

\newcommand{\thmref}[1]{Theorem~\ref{#1}}
\newcommand{\secref}[1]{\S\ref{#1}}
\newcommand{\lemref}[1]{Lemma~\ref{#1}}
\newcommand{\propref}[1]{Proposition~\ref{#1}}
\newcommand{\nc}{\newcommand}
\nc{\on}{\operatorname}
\nc{\ch}{\on{ch}}
\nc{\Z}{{\Bbb Z}}
\nc{\C}{{\Bbb C}}
\nc{\Pro}{{\Bbb P}}
\nc{\R}{{\Bbb R}}
\nc{\cond}{|\,}
\nc{\bib}{\bibitem}
\nc{\pone}{\Pro^1}
\nc{\Li}{{\cal L}}
\nc{\La}{{\cal A}}
\nc{\pa}{\frac{\partial}{\partial z}}
\nc{\bu}{{\bold u}}
\nc{\bv}{{\bold v}}
\nc{\bw}{{\bold w}}
\nc{\mm}{{\cal M}}
\nc{\mc}{{\cal M}}
\nc{\mon}{\text{mon}}

\title[Coinvariants of subalgebras of the Virasoro algebra]{Coinvariants of
nilpotent subalgebras of the Virasoro algebra and partition identities}

\author{Boris Feigin}
\address{Landau Institute for Theoretical Physics, Moscow 117334, Russia}

\author{Edward Frenkel}
\address{Department of Mathematics \\ Harvard University \\ Cambridge, MA
02138, USA}
\thanks{Research of the second author was supported by a Junior Fellowship
from the Harvard Society of Fellows and by NSF grant DMS-9205303}

\dedicatory{Dedicated to I.M. Gelfand on his 80th birthday}

\maketitle

\section{Introduction}

Let $V^{m,n}_{p,q}$ be the irreducible representation of the Virasoro
algebra $\Li$ with central charge $c_{p,q}=1-6(p-q)^2/pq$ and highest
weight $h_{m,n}=[(np-mq)^2-(p-q)^2]/4pq$, where $p,q>1$ are relatively
prime integers, and $m,n$ are integers, such that $0<m<p, 0<n<q$. For
fixed $p$ and $q$ the representations $V^{m,n}_{p,q}$ form the $(p,q)$
minimal model of the Virasoro algebra \cite{bpz}.

For $N>0$ let $\Li_N$ be the Lie subalgebra of the Virasoro algebra,
generated by $L_i, i<-N$. There is a map from the Virasoro algebra to the
Lie algebra of polynomial vector fields on $\C^*$, which takes $L_i$ to
$z^{-i+1} \pa$, where $z$ is a coordinate. This map identifies $\Li_N$ with
the Lie algebra of vector fields on $\C$, which vanish at the origin along
with the first $N+1$ derivatives. The Lie algebra $\Li_{2N}$ has a family
of deformations $\Li(p_1,...,p_{N+1})$, which consist of vector fields,
vanishing at the points $p_i \in \C$ along with the first derivative.

For a Lie algebra $g$ and a $g-$module $M$ we denote by $H(g,M)$ the space
of coinvariants (or 0th homology) of $g$ in $M$, which is
the quotient $M/g\cdot M$, where $g\cdot M$ is the subspace of $M$,
linearly spanned by vectors $a\cdot x, a \in g, x \in M$. We will prove the
following result.

\begin{thm} For any irreducible representation $V$ of the $(2,2r+1)$
minimal model
$$\dim H(\Li_{2N},V) = \dim H(\Li(p_1,...,p_{N+1}),V).$$
\end{thm}

This statement was proved in \cite{fefu} for $N=2$ and irreducible
representations of general minimal models.

As an application, we will give a new proof of the Gordon identities, which
relies on two different computations of the characters of the irreducible
representations of the $(2,2r+1)$ minimal models.

Let us note that {\em a priori} $\dim H(\Li_{2N},V) \geq \dim
H(\Li(p_1,\ldots,p_{N+1}),V)$. Indeed, if we have a family of Lie algebras,
then the dimension of homology is the same for generic points of the space
of parameters of the family, but it may increase at special points.

Each minimal model of the Virasoro algebra associates to a punctured
complex curve, with a representation inserted at each puncture, a linear
space, which is called the space of conformal blocks. This space can be
defined as the space of coinvariants of the Lie algebra of meromorphic
vector fields on this curve, which are allowed to have poles only at the
punctures, in the tensor product of these representations
(cf. \cite{bpz}, mathematical aspects of this correspondence are treated in
detail in \cite{fefu}, \cite{bfm}).

In this language, the space $H(\Li(p_1,\ldots,p_{N+1}),V^{m,n}_{p,q})$ is
isomorphic to the direct sum of the spaces of conformal blocks, associated
to the projective line with punctures $p_1,\ldots,p_{N+1}$ with all
possible insertions from the $(p,q)$ minimal model, and $\infty$ with the
insertion of $V^{m,n}_{p,q}$. The dimension of this space can be calculated
by a Verlinde type argument by deforming our curve -- the projective line
with $N+2$ punctures -- to a joint union of $N$ projective lines, each
with 3 punctures. Under this deformation the dimension of the homology does
not change \cite{fefu},
\cite{bfm}, and can therefore be reduced to the well-known result on the
dimensions of the spaces of conformal blocks, associated to the projective
line with 3 insertions -- the fusion coefficients.

This gives an explicit formula for $\dim
H(\Li(p_1,\ldots,p_{N+1}),V^{m,n}_{p,q})$, which represents a lower bound
for $\dim H(\Li_{2N},V^{m,n}_{p,q})$. We can also obtain an upper bound by
a different method.

This method is based on the calculation of the annihilating ideal of a
minimal model \cite{feno}. Let us recall the definition of the annihilating
ideal. For $c \in \C$ let $M_c$ be the vacuum Verma module with central
charge $c$. It is generated by a vector $v_c$, such that $L_i v_c = 0, i
\geq -1$. This representation has the structure of a vertex operator
algebra (VOA) \cite{fzhu}, which acts on any irreducible module $M$ of the
Virasoro algebra with the same central charge. It means that each vector of
$M_c$ defines a local current (or field), which is a formal power series in
$z$ and $z^{-1}$, whose Fourier coefficients are linear operators, acting
on $M$. These local currents can be constructed as follows. Consider the
projective line $\C\pone$ with three punctures: $0$ and $\infty$ with the
insertions of $M$, and $z$ with the insertion of $M_c$. The space of
conformal blocks is one-dimensional in this case. In other words, the space
of coinvariants of the Lie algebra $\La(0,z)$ of meromorphic vector fields
on $\C\pone$, which are allowed to have poles only at $0,
\infty$ and $z$, in $M \otimes M \otimes M_c$, is
one-dimensional. It has a canonical generator -- the projection of the
tensor product of the highest weight vectors. The dual to this generator
defines for each vector of $M_c$ a linear operator from $M$ to $M$,
depending on $z$, which is our local current.

These local currents can be constructed explicitly. The monomials
$$\{L_{-m_1}\dots L_{-m_l}v_c\cond m_1\geq m_2\geq\dots \geq m_l>1\}$$
linearly span $M_c$. The corresponding local currents are equal to
\begin{equation}   \label{current}
\frac{1}{(m_1-2)!}\ldots\frac{1}{(m_l-2)!} :\partial^{m_1-2}_z T(z)\ldots
\partial^{m_l-2}_z T(z):,
\end{equation}
where $T(z) = \sum_{i \in \Z} L_i z^{-i-2}$.
The Fourier components of thus defined local currents form a Lie algebra
$U_c(\Li)_{loc}$, which is called the local completion of the universal
enveloping algebra of the Virasoro algebra with central charge $c$
\cite{fefr},\cite{feno}. They act on any representation of $\Li$, which has
the same central charge.

The module $M_c$ is irreducible, if and only if $c$ is not equal to
$c_{p,q}$. However, if $c=c_{p,q}$, then $M_c$ contains a (unique) singular
vector. The quotient of $M_{c_{p,q}}$ by the submodule, generated by this
singular vector, is isomorphic to $V^{1,1}_{p,q}$. This representation,
which is called the vacuum representation of the $(p,q)$ minimal model,
defines another VOA, which is the quotient of the VOA of $M_{c_{p,q}}$ in
the appropriate sense \cite{fhl}. An irreducible representation $M$ of the
Virasoro algebra with central charge $c_{p,q}$ is a module over this VOA
\cite{fhl}, if and only if the space of coinvariants of the Lie algebra
$\La(0,z)$ in $M \otimes M \otimes V^{1,1}_{p,q}$ is
one-dimensional. An explicit computation of the dimension of this space,
which was made in \cite{fefu}, tells us that it is so, if and only if $M$
is an irreducible representation of the $(p,q)$ minimal model.

The Fourier components of the local current, corresponding to the singular
vector of $M_{c_{p,q}}$, generate an ideal in $U_{c_{p,q}}(\Li)_{loc}$,
which is called the annihilating ideal of the $(p,q)$ minimal model. Any
element of this ideal acts trivially on any representation of the $(p,q)$
minimal model.

Imposing this condition with respect to the generators of the annihilating
ideal immediately leads to certain linear relations among the monomial
elements of $V^{m,n}_{p,q}$. This allows to estimate the dimension of the
coinvariants $H_0(\Li_{2N},V^{m,n}_{p,q})$ from above.

The peculiarity of the $(2,2r+1)$ models is that this upper bound coincides
with the lower bound, and is therefore exact. This proves Theorem 1 and
leads to a nice combinatorial description of the minimal representations of
these models. Namely, as a $\Z-$graded linear space, such a representation
is isomorphic to the quotient of the space of polynomials in infinitely
many variables by a certain monomial ideal. Thus we obtain an expression
for the character of this module, which coincides with the right hand side
of one of the Gordon identities. But it is known that this character is
equal to the left hand side of the identity. Hence, we obtain a new proof
of the Gordon identities.

Implementing this program for other minimal models would lead to a result,
analogous to Theorem 1, as well as to nice character formulas for
irreducible representations. However, for general minimal models, for which
the representations are ``smaller'' and the structure of the annihilating
ideal is more complicated, one should impose extra conditions,
corresponding to other elements of the annihilating ideal, to obtain the
exact upper bound.

\section{Lower bound: genus 0 conformal blocks}

In this section we will calculate the dimension of
$H(\Li(p_1,\ldots,p_{N+1}),V)$ for an irreducible representation $V$ of a
$(p,q)$ minimal model.

Denote by $\La(p_1,\ldots,p_{N+1})$ the Lie algebra of meromorphic vector
fields on $\C\pone$, which are allowed to have poles only at the distinct
points $p_1,\ldots,p_{N+1} \in \C$, and $\infty$. There is an embedding of
this Lie algebra into the direct sum of $N+2$ Virasoro algebras, and thus
it acts on tensor products $V_1 \otimes \ldots \otimes V_{N+2}$ of $N+2$
reprersentations of the Virasoro algebra with the same central charge.

Let us fix central charge $c_{p,q}$ and an irreducible representation $V =
V_{N+2}$ of the $(p,q)$ minimal model. The following statement follows from
the general results on computation of conformal blocks, outlined in Section
5 of \cite{fefu} (cf. \cite{bfm} for details).

\begin{prop}
(1) $\dim H(\Li(p_1,\ldots,p_{N+1}),V) = \sum \dim
H(\La(p_1,\ldots,p_{N+1}),V_1 \otimes \ldots \otimes V_{N+1} \otimes V)$,
where the sum is taken over all $(N+1)-$tuples $V_1,\ldots,V_{N+1}$ of
irreducible representations of the $(p,q)$ minimal model.

(2) $\dim H(\La(p_1,\ldots,p_{N+1}),V_1 \otimes \ldots \otimes V_{N+1}
\otimes V) = \sum \dim H(\La(p_1,\ldots,p_N),V_1 \otimes \ldots \otimes V_N
\otimes W) \cdot \dim H(\La(0,1),W \otimes V_{N+1} \otimes V),$ where the
sum is taken over all irreducible representations $W$ of the $(p,q)$
minimal model.
\end{prop}

Using this result, we can calculate $\dim H(\Li(p_1,\ldots,p_{N+1}),V)$ by
induction.

Let $V_1,\ldots,V_s$ be the set of all irreducible representations of the
$(p,q)$ minimal model ($s=(p-1)(q-1)/2$), and let $\bu_N$ be the
$s-$vector, whose components are $u^i_N = \dim
H(\Li(p_1,\ldots,p_{N+1}),V_i), i=1,\ldots,s$. Put $c_{ijk} = \dim
H(\La(0,1),V_i \otimes V_j
\otimes V_k)$. The numbers $c_{ijk}$ are usually called the fusion
coefficients. They define the fusion algebra of this minimal model, which
has generators $g_i, 1 \leq i \leq s$ and relations $g_i \cdot g_j = \sum_k
c_{ijk} g_k$.

Let us introduce the $s \times s$ matrix $\mm = \mm_{p,q}$, whose $(i,j)$th
entry is equal to $\sum_k c_{ijk}$. This matrix can be interpreted as the
matrix of action of the sum $\sum_k g_k$ on the fusion algebra.

According to Proposition 2, $u^i_N = \sum_{j,k} c_{ijk} u^j_{N-1}$.
This gives us the recursion relation for $\bu_N$:
$$\bu_N = \mm \bu_{N-1}.$$

For any $i$, the representation $V_i$ is generated from the highest weight
vector $v_i$ by the action of the Lie algebra $\Li_0$, hence $u^i_0 = 1$.
Therefore, we obtain the formula $$\bu_N = \mm^N \bu_0,$$ where $\bu^t_0 =
[1,1,\ldots,1]$.

This formula enables us to calculate the dimensions $u^i_N$ explicitly by
diagonalizing the matrix $\mm$.

In the rest of the paper we will focus on the $(2,2r+1)$ models. In such a
model we have $r$ irreducible representations $V_i = V^{1,i}_{2,2r+1},
i=1,\ldots,r$. The $(i,j)$th entry of the corresponding $r \times r$
matrix $\mm$ is equal to min$\{i,j\}$. This matrix is equal to the square
of the matrix
\begin{equation}   \label{matrix}
M_r(x)=
\begin{pmatrix}
0 & 0 & \hdots & 0 & 1 \\
0 & 0 & \hdots & 1 & 1 \\
\hdotsfor{5} \\
0 & 1 & \hdots & 1 & 1 \\
1 & 1 & \hdots & 1 & 1
\end{pmatrix}.
\end{equation}

This fact has the following interpretation: in the fusion algebra of this
model $\sum_k g_k$ is equal to the square of $g_r$, and the action of $g_r$
is given by matrix $M_r$.

Let us introduce vectors ${\bold x}_i$, whose $i$th component is equal to
1, and all other component are equal to 0. Since $\bu_0 = M_r^2 {\bold
x}_1$, we have $u^i_N = {\bold x}^t_i M_r^{2N+2} {\bold x}_1$. This gives
us a lower bound for $v^i_N = \dim H(\Li_{2N},V_i)$.

\begin{prop} $v^i_N \geq {\bold x}^t_i M_r^{2N+2} {\bold x}_1$.
\end{prop}

In the next Section we will show that this bound is exact.

\section{Upper bound: the annihilating ideal}

The annihilating ideal of the $(2,2r+1)$ minimal model is the ideal of the
Lie algebra $U_{c_{2,2r+1}}(\Li)_{loc}$, generated by the Fourier
components of the local current, corresponding to the singular vector of
the vacuum Verma module $M_{c_{2,2r+1}}$. All elements of this ideal act by
0 on any irreducible representation of the $(2,2r+1)$ minimal model. In
particular, the generators of the ideal act by 0. This leads to certain
relations between vectors in the irreducible minimal representations, as
explained in
\cite{feno}.

Introduce a filtration on the irreducible representation $V_i$ of the
$(2,2r+1)$ models as follows: $0 \subset V^0_i \subset V^1_i \subset
\ldots \subset V^\infty_i = V_i,$ where $V^k_i$ is linearly spanned by the
monomials $L_{-m_1} \dots L_{-m_l} v_i$, such that $m_1\geq m_2\geq\dots
\geq m_l\geq 1$, and $l \leq k$. Let $\Omega_i$ be the adjoint graded space:
$$\Omega_i = \oplus_{k \geq 0}
\Omega^k_i,\quad \Omega^k_i = V^k_i/V^{k-1}_i.$$ The space $\Omega_i$ is a
polynomial algebra in variables $a_j, j>0$, where $a_j$ is the image of
$L_{-j} v_i$.

The symbol of the singular vector of $M_{c_{2,2r+1}}$ is equal to $L_{-2}^r
v_0$. According to formula \eqref{current}, the symbol of the corresponding
local current is equal to $:T(z)^r:$. Hence the symbols of the Fourier
components of this current are given by the formula
$$\sum_{j_1+\ldots +j_r=n} :L_{-j_1}\ldots L_{-j_r}:.$$
Their action on $\Omega_i$ is given by the multiplication by
\begin{equation}   \label{Sn}
S_n = \sum_{j_k>0; j_1+\ldots +j_r=n} a_{j_1}\ldots a_{j_r},
\end{equation}
for $n\geq r$, and 0 for $n<r$.

It is known that $V_i$ is the quotient of the Verma module
with highest weight $h_{1,i}$ by the maximal submodule, generated by two
singular vectors. The symbol of one of them is equal to $L_{-1}^i v_i$.
This gives us a surjective map from the quotient $\Omega'_i$ of
$\C[a_j]_{j>0}$ by the ideal $I_i$ generated by $S_n, n>r$, and $a_1^i$, to
$\Omega_i$.

Let $\Omega^{\mon}_i$ be the quotient of $\C[a_j]_{j>0}$ by the ideal
$I_i^{\mon}$, generated by the monomials $a_j^u a_{j+1}^v, 0\leq u<r,
u+v=r, j>0$, and $a_1^i$.

We can introduce a $\Z-$grading on the module $V_i$ by putting $\deg v_i =
0, \deg L_{-j} = j$. The space $\Omega_i$ inherits this grading. We can
also introduce a compatible $\Z-$grading on the space $\C[a_j]_{j>0}$ by
putting $\deg a_{-j} = j$. The spaces $\Omega'_i$ and $\Omega^{\mon}_i$
inherit this grading.

For any $\Z-$graded linear space $V =
\oplus_{n\geq 0} V(n)$, such that $\dim V(n)\leq \infty$, let $\ch V =
\sum_{n\geq 0} \dim V(n) q^n$ be its character. We will write $\ch V \leq \ch
V'$, if $\dim V(n) \leq \dim V'(n)$ for any $n\geq 0$.

We have $$\ch V_i = \ch \Omega_i \leq \ch \Omega'_i \leq \ch
\Omega^{\mon}_i.$$ The last inequality follows from the fact that each of
the generators $S_n$ of the ideal $I_i$ has as a summand one and only one
generator of the ideal $I_i^{\mon}$, namely, $a_j^u a_{j+1}^v$ is a summand
of $S_{rj+v}$.

Now, by the construction of our filtration, the character of coinvariants
$H(\Li_{2N},V_i)$ is equal to the character of the quotient $\Omega_{i,N}$
of the space $\Omega_i$ by the ideal, generated by $a_j$ with $j>2N$.
Again, we have:
\begin{equation}   \label{inequality}
\ch H(\Li_{2N},V_i) = \ch \Omega_{i,N} \leq \ch \Omega^{\mon}_{i,N},
\end{equation}
where $\Omega^{\mon}_{i,N}$ is the quotient of $\Omega^{\mon}_i$ by the
ideal, generated by $a_j, j>2N$. Therefore,$$v^i_N = \dim H(\Li_{2N},V_i)
\leq \omega^i_N = \dim \Omega^{\mon}_{i,2N}.$$ We can easily
calculate $\omega^i_N$ by induction \cite{frsz}.

Let $C^r_i = \{(m_1,\dots,m_l)\cond m_1\geq\dots \geq m_l\geq 1, m_i\geq
m_{i+r-1}+2, m_{l-i+1}>1\}$. The monomials $$\{ a_{m_1}\dots a_{m_l}\cond
(m_1,\dots,m_l)\in C^r_i\}$$ constitute a linear basis in $\Omega^{\mon}_i$.

For every integer $N>0$, introduce the subspaces $W^i_{k,N}, 1\leq k\leq
r$, of $\Omega^{\mon}_i$, which are linearly spanned by the monomials $$\{
a_{m_1}\dots a_{m_l}\cond (m_1,\dots,m_l)\in C^r_i, m_1\leq N, m_{k} \leq
N-1\}.$$ Clearly, $W^i_{1,N}$ is isomorphic to $\Omega^{\mon}_{i,N-1}$.

The dimension of the space $W^i_{k,N}$, which is spanned by our monomials,
in which $a_N$ is allowed in the power less than $k$, is equal to the sum
of dimensions of the spaces $W^i_{l,N-1}$ with $l = 1,\ldots,r-k+1$,
because we have the relations $a_{N-1}^u a_N^v = 0$ for $u+v=r$. Let us
introduce the $r-$vector $\bw^i_N$, whose components are $w^i_{k,N} = \dim
W^i_{k,N}, k=1,\ldots,r$. We obtain the formula \cite{frsz}: $$\bw^i_N = M_r
\bw^i_{N-1},$$ where $M_r$ is given by \eqref{matrix}, which shows that
$\bw^i_{2N+1} = M_r^{2N} \bw^i_1$.

\begin{rem} As shown in \cite{frsz}, there is a $q-$deformation of this
formula, which gives an expression for the characters of $W^i_{k,N}$.
\end{rem}

By definition, $w^i_{k,1} = \text{min}\{i,k\}$. One can check that
$\bw^i_1 = M_r^2 {\bold x}_i$, and so $\bw^i_{2N+1} = M_r^{2N+2} {\bold
x}_i$. Therefore, $\omega^i_N = w^i_{1,2N+1} = {\bold x}_1^t M_r^{2N+2}
{\bold x}_i$, and this gives us an upper bound for $v^i_N$.

\begin{prop}
$$v^i_N \leq {\bold x}_1^t M_r^{2N+2} {\bold x}_i.$$
\end{prop}

But since $M_r^t = M_r$, this upper bound coincides with the lower bound
from Proposition 3. Therefore, we have the equality $$u^i_N = v^i_N =
\omega^i_N = {\bold x}_1^t M_r^{2N+2} {\bold x}_i.$$ This completes the
proof of Theorem 1.

Note that we have also proved that $\Omega_i$ is isomorphic to $\Omega'_i =
\C[a_j]_{j>0}/I_i$.

\begin{rem} For general minimal models the graded space $\Omega$ of an
irreducible representation is also isomorphic to the quotient of
$\C[a_j]_{j>0}$ by a certain ideal $I$. This ideal contains the operators
$S_n$, given by formula \eqref{Sn} with $r=(p-1)(q-1)/2$. They correspond
to the action of the symbols of generators of the annihilating ideal of the
minimal model. In general, however, these elements do not generate the
ideal $I$. There are other generators, corresponding to the action of the
symbols of other elements of the annihilating ideal on $\Omega$. It is an
interesting problem to find explicit formulas for them.  This will
hopefully lead to a nice combinatorial description of general irreducible
representations.
\end{rem}

\section{Application: Gordon identities}

By formula \eqref{inequality}, $\dim H(\Li_{2N},V_i)(n) \leq \dim
\Omega^{\mon}_{i,2N}(n)$ for any $n$ and $N$. In the previous Section we
proved that $\dim H(\Li_{2N},V_i) = \dim \Omega^{\mon}_{i,2N}$, therefore
$\dim H(\Li_{2N},V_i)(n) = \dim \Omega^{\mon}_{i,2N}(n)$ for any $n$ and
$N$. But, clearly, $\dim H(\Li_{2N},V_i)(n) =
\dim V_i(n)$ and $\dim \Omega^{\mon}_{i,2N}(n) = \dim \Omega^{\mon}_i(n)$
for $N$ large enough. Hence $\dim V_i(n) = \dim \Omega^{\mon}_i(n)$ for any
$n$, and $\ch V_i = \ch \Omega^{\mon}_i$. This can be interpreted as
follows.

\begin{prop} The monomials
$$\{ L_{-m_1}\dots L_{-m_l}v_i \cond (m_1,\dots,m_l)\in C^r_i\}$$
constitute a linear basis in the irreducible representation $V_i$ of the
$(2,2r+1)$ minimal model.
\end{prop}

Thus, we obtain the following formula for the character of the module
$V_i$:
\begin{equation}   \label{character}
\ch V_i = \sum_{n\geq 0} |C^r_i(n)| q^n,
\end{equation}
where $C^r_i(n)$ is the subset of $C^r_i$, which consists of the elements
$(m_1,\ldots,m_l)
\in C^r_i$, for which $\sum m_j = n$. The right hand side of the formula
\eqref{character} is known to be equal to $$\sum_{n_1,\ldots,n_{r-1}\geq 0}
\frac{q^{N_1^2+\ldots +N_{r-1}^2+N_i+\ldots +N_{r-1}}}{(q)_{n_1}\ldots
(q)_{n_{r-1}}},$$ where $N_j=n_j+\ldots +n_{r-1}$, and $(q)_n =
\prod_{l=1}^{l=n}(1-q^l)$ (cf. \cite{andr}).

On the other hand, it is known \cite{kacw,kac} that $$\ch V_i =
\prod_{n>0,n\neq 0,\pm i \text{mod}(2r+1)} (1-q^n)^{-1}.$$ This formula
follows from the Weyl-Kac type character formula for irreducible minimal
representations of the Virasoro algebra.

Thus, we have obtained a new proof of the Gordon identities:
$$\prod_{n>0,n\neq 0,\pm i \text{mod}(2r+1)} (1-q^n)^{-1} =
\sum_{n_1,\ldots,n_{r-1}\geq 0} \frac{q^{N_1^2+\ldots
+N_{r-1}^2+N_i+\ldots +N_{r-1}}}{(q)_{n_1}\ldots (q)_{n_{r-1}}}.$$

For $r=2$ these are the Rogers-Ramanujan identities:
$$\prod_{n>0,n\neq 0,\pm 1 \text{mod} 5} (1-q^n)^{-1} =
\sum_{n\geq 0} \frac{q^{n(n+1)}}{(q)_n}\quad\text{and}\quad
\prod_{n>0,n\neq 0,\pm 2 \text{mod} 5} (1-q^n)^{-1} =
\sum_{n\geq 0} \frac{q^{n^2}}{(q)_n}.$$
They correspond to two irreducible representations of the $(2,5)$ minimal
model.

\vspace{5mm}
\noindent{\bf Acknowledgements} The main part of this work was done while
B.F. was visiting the Isaac Newton Institute of the University of
Cambridge. He would like to the thank the Institute for hospitality. E.F.
thanks A.Szenes for valuable discussions.

\vspace{5mm}
\bibliographystyle{amsplain}

\end{document}